# Small Fermi pockets intertwined with charge stripes and pair density wave order in a kagome superconductor

Hong Li[1,*], Dongjin Oh[2,*], Mingu Kang[2,*], He Zhao[1], Brenden R Ortiz[3], Yuzki Oey[3], Shiang Fang[2], Zheng Ren[1], Chris Jozwiak[4], Aaron Bostwick[4], Eli Rotenberg[4], Joseph G. Checkelsky[2], Ziqiang Wang[1], Stephen D. Wilson[3], Riccardo Comin[2,§] and Ilija Zeljkovic[1,§]

[1]Department of Physics, Boston College, Chestnut Hill, MA 02467, USA
[2]Department of Physics, Massachusetts Institute of Technology, Cambridge, Massachusetts 02139, USA
[3]Materials Department, University of California Santa Barbara, Santa Barbara, California 93106, USA
[4]Advanced Light Source, E. O. Lawrence Berkeley National Laboratory, Berkeley, California 94720, USA.
*These authors contributed equally.
§Corresponding authors: rcomin@mit.edu and ilija.zeljkovic@bc.edu

**The kagome superconductor family $AV_3Sb_5$ ($A$=Cs, K, Rb) emerged as an exciting platform to study exotic Fermi surface instabilities. Here we use spectroscopic-imaging scanning tunneling microscopy (SI-STM) and angle-resolved photoemission spectroscopy (ARPES) to reveal how the surprising cascade of higher and lower-dimensional density waves in $CsV_3Sb_5$ is intimately tied to a set of small reconstructed Fermi pockets. ARPES measurements visualize the formation of these pockets generated by a 3D charge density wave transition. The pockets are connected by dispersive $q^*$ wave vectors observed in Fourier transforms of STM differential conductance maps. As the additional 1D charge order emerges at a lower temperature, $q^*$ wave vectors become substantially renormalized, signaling further reconstruction of the Fermi pockets. Remarkably, in the superconducting state, the superconducting gap modulations give rise to an in-plane Cooper pair-density-wave at the same $q^*$ wave vectors. Our work demonstrates the intrinsic origin of the charge-stripes and the pair-density-wave in $CsV_3Sb_5$ and their relationship to the Fermi pockets. These experiments uncover a unique scenario of how Fermi pockets generated by a parent charge density wave state can provide a favorable platform for the emergence of additional density waves.**

Correlated electron systems are often characterized by coexisting electronic orders vying for the same electronic states at the Fermi level. As a result, the residual Fermi surface exhibits smaller Fermi surfaces, or "pockets", that govern many aspects of the low-energy physics. Identifying the existence of such pockets and understanding their renormalization driven by the emergent orders has been of immense interest in many complex solid state systems. In cuprate superconductors for example, there has been a long standing debate whether Fermi pockets existed, in an attempt to understand their connection to the correlated liquid states emerging upon doping the parent Mott insulating phase [1–4].

Kagome superconductors in the $AV_3Sb_5$ [5–8] family recently emerged as a new materials platform to theoretically [9–16] and experimentally [17,18,27–30,19–26] study exotic correlated and topological quantum states intertwined with Fermi surface instabilities. The rich array of electronic phenomena observed includes an unusual number of density waves with varying morphology and dimensionality, which onset at different temperatures upon cooling down and all co-exist with superconductivity at low temperature. This includes the 3D charge density wave (CDW) with a $2a_0$ by $2a_0$ period in the kagome plane (onset temperature $T^* \sim$ 70-100 K [6,8,21,26,31,32]), a $4a_0$ charge-stripe order ($T_{4a0} \sim$ 50-60 K [18]) and a $4a_0/3$ by $4a_0/3$ Cooper pair-density wave ($T_c \sim$ 3-4 K [19]). The formation of the 3D CDW at $T^*$ has now been well-established and explored by a variety of bulk probes [21,26,33,34], and its $2a_0$ in-plane wave-length has been expected to naturally arise on a hexagonal lattice [35]. In contrast, the surprising emergence of lower-dimensional density waves – the 1D $4a_0$ charge-stripes [18] and the in-plane pair-density wave [19] – has been difficult to capture by the same bulk-sensitive tools. This prompted an intense debate in the community on whether these states are intrinsic to the kagome lattice or have a different origin, e.g. a surface reconstruction [36,37]. Recent theory [15] has predicted that small Fermi pockets generated by the high-temperature CDW transition could be instrumental in the subsequent creation of exotic density waves at lower temperatures. In principle, this would lead to distinct fingerprints in the bulk electronic structure of the kagome bands associated with the transitions, but this is yet to be uncovered.

Here we use a combination of angle-resolved photoemission spectroscopy (ARPES) and spectroscopic-imaging scanning tunneling microscopy (SI-STM) to unveil an intimate connection between density waves and reconstructed Fermi pockets in $A$V$_3$Sb$_5$. Our ARPES measurements reveal six small ellipsoidal hole-like pockets generated by the 2 x 2 CDW state forming at $T^*$. Scattering between these pockets leads to new, dispersive wave vectors **q*** in SI-STM measurements, oriented along each reciprocal lattice direction. While the three **q*** peaks all lie along the Γ-K directions of the original Brillouin zone when the $4a_0$ charge-stripe order is absent, we discover that the morphology of these scattering peaks changes dramatically when the charge-stripe order sets in. Namely, the dispersive nature of one **q*** peak becomes markedly suppressed, while the other two **q*** peaks remain dispersive and exhibit a slight deviation from the high-symmetry axes. This strongly suggests further reconstruction of the pockets as the charge-stripe order emerges, which could explain some of the smaller frequencies in quantum oscillations experiments of CsV$_3$Sb$_5$ [33,34,38–41]. Remarkably, the Cooper pair-density wave [19] that condenses in the superconducting state emerges at the same **q*** wave vectors that connects the hole pockets in reciprocal space. Our experiments reveal a direct link between vanadium kagome orbital derived Fermi pockets, an inherent feature of the bulk electronic band structure, and the surprising cascade of lower-dimensional density waves in $A$V$_3$Sb$_5$.

Bulk single crystals of $A$V$_3$Sb$_5$ exhibit a layered structure consisting of V$_3$Sb$_5$ layers stacked between $A$-site alkali metal layers (Fig. 1a). The crystals tend to cleave between the $A$-site layer and the Sb layer (Methods), resulting in two different types of surfaces: $A$ termination and the Sb termination [17–20,28,42]. In STM experiments, we focus on the Sb surface positioned directly above the kagome plane due to its structural stability and the direct access to bulk vanadium-derived kagome bands [18]. Similar to previous experiments [18–20,42], STM topographs of the Sb surface of CsV$_3$Sb$_5$ show a unidirectional electronic modulation related to the $4a_0$ charge-stripe order (Fig. 1b,e), which forms below about 50-60 K [18]. As more clearly seen from the Fourier transform of the STM topograph at low temperature (Fig. 1b,e), the $4a_0$ charge order spatially co-exists with the $2a_0$ by $2a_0$ charge-density wave [6,32,43].

By replacing Cs with K in an identical $A$V$_3$Sb$_5$ crystal structure, the long-range $4a_0$ charge-stripe order vanishes and cannot be detected on the equivalent Sb surface termination at 4.5 K (Fig. 1c,f) [17,28], while most of the other known properties remain qualitatively the same. For example, superconductivity in both materials emerges from the same metallic normal state with a $2a_0$ by $2a_0$ CDW in the kagome plane, which also breaks rotational symmetry of the lattice [17,28,44–46]. In STM measurements, this high-temperature rotation symmetry breaking can be visualized by anisotropic CDW amplitudes [28,44,45], with one preferred direction being markedly different from the other two that are nearly indistinguishable (insets in Supplementary Figure 1b,e). This symmetry breaking gives rise to three types of domains rotated by 120 degrees from one another observed by optical birefringence measurements [46]. Muon spin spectroscopy [47,48], magneto-optical Kerr measurements [46] and circular dichroism [46] have also revealed signatures of time-reversal symmetry breaking in both $CsV_3Sb_5$ and $KV_3Sb_5$. The two materials provide an exciting playground for the exploration of Fermi surface reconstruction driven by emergent density waves, and a fortuitous opportunity to use $KV_3Sb_5$ as a foil for comparison with $CsV_3Sb_5$ to understand the emergence of the charge-stripe order.

We first measure the temperature-dependent Fermi surface and energy-momentum dispersions of $AV_3Sb_5$ using ARPES. While previous experiments investigated the overall renormalization of the Fermi surface driven by the 2 x 2 CDW transition [22,49–52], here we specifically focus on the formation of small Fermi pockets to uncover their renormalization in connection to the emergent density waves using a combination of ARPES and SI-STM. Our ARPES measurements clearly reveal the formation of Fermi pockets, which can be seen in the second Brillouin zone in all $AV_3Sb_5$ systems studied here: $KV_3Sb_5$ (Fig. 2a), $CsV_3Sb_5$ and $RbV_3Sb_5$ (Supplementary Figure 2). Notably, these pockets appear at the $M_2$ points of the reduced Brillouin zone (blue dashed hexagons in Fig. 2a) induced by CDW order, and they disappear in the normal state above the CDW transition (Fig. 2b).

Further insights into the formation of small pockets can be obtained by examining the detailed energy-momentum dispersions. Consistent with the previous reports [22,49–52], the $p$-type van Hove singularities (vHSs) from the K1 and K2 bands are clearly visible in the ARPES spectra above the CDW transition (Fig. 2h,i). In the CDW state, the main CDW gap of ~80 meV opens at the vHS of the K1 band, forming the 'M'-shaped dispersion along the $\overline{K}$-$\overline{M}$-$\overline{K}$ direction (Fig. 2j) [22,53]. Importantly, this opening of the CDW gap at the M point drives the reconstruction of the K1 Fermi surface around the $M_2$ point, resulting in the observed small Fermi pockets (Fig. 2c,e) [52]. The pockets exhibit hole-like dispersion along both axes of the ellipse (Fig. 2e,m). We note that along the minor axis, the hole-like dispersion is a consequence of the back-folding of the K1 band across the $M_2$ point (Fig. 2g,m) with the reduced spectral weight of the folded side (see Fig. 2a,c). In addition, the K2 and K2′ bands show clear back-banding in the CDW state, suggesting the existence of a CDW gap (red and blue arrows in Fig. 2m). The overall spectral weight of the observed Fermi surface and dispersions are closely reproduced by DFT calculations (Fig. 2f,g, Methods) [54]. We note that while earlier work reported evidence suggesting the existence of Fermi pockets in $KV_3Sb_5$ [52], our combination of ARPES and DFT unambiguously demonstrates the formation of small Fermi hole pockets in the CDW state of all $AV_3Sb_5$ systems, arising from the interplay of vHS and the CDW gap. Moreover, the estimated area of the elliptical pockets (obtained as $\pi \cdot k_{Fa} \cdot k_{Fb}$, with $k_{Fa}$ and $k_{Fb}$ being the major and the minor radius of the pocket, respectively) found in the present study translates to a quantum oscillation frequency of ~86.8 ± 26.2 T via the Onsager relation, in agreement with recent quantum oscillation studies within the experimental resolution (Supplementary Table 1). The quantitative correspondence between the ARPES and quantum oscillation experiments further confirms the bulk nature of the observed Fermi pockets.

Complementary to ARPES measurements of the electronic band structure in the normal state, we image the scattering and interference of electrons using SI-STM. Fourier transforms of d$I$/d$V$(**r**, $V$) maps on the Sb surface of the two systems display similar, dispersive scattering wave vectors **q**$_1$ and **q**$_2$ (Supplementary Figure 1 and 3). In addition to these previously reported wave vectors, our high-resolution SI-STM measurements reveal a set of new, dispersive scattering wave vectors **q**$^*_i$, where $i = a, b$ or $c$ lattice direction (Fig. 3a). For simplicity, we first examine **q**$^*_i$ in KV$_3$Sb$_5$, in the absence of $4a_0$ charge-stripe ordering. In contrast to **q**$_2$ scattering wave vectors that are markedly unidirectional (Supplementary Figure 1), **q**$^*_i$ wave vectors appear along all three atomic Bragg peak **Q**$^i_{Bragg}$ directions (Fig. 3a,b). They are detectable around Fermi level and disperse with energy in a similar manner along the three lattice directions (Fig. 3d-f, Supplementary Figure 4). The dispersive nature of **q**$^*_i$ suggests that these wave vectors originate from scattering between different points on the constant energy contour, which changes concomitant with the band structure evolution. The magnitude and the direction of the dispersion of **q**$^*_i$ from SI-STM measurements are beautifully consistent with the pockets extracted from ARPES (Supplementary Figure 5). The scattering primarily occurs between the outer sides of the pockets with the significantly larger spectral weight as observed in ARPES measurements (Supplementary Figure 5, 6). It is important to note that these electronic states correspond to residual electronic states near Fermi level after band folding and gapping induced by the 2 x 2 CDW state in the kagome plane. Taken together, our data demonstrates an intimate relationship between the emergence of **q**$^*_i$ and the existence of Fermi pockets near the reduced Brillouin zone boundary (Fig. 3c).

Interestingly, the morphology of **q**$^*_i$ changes profoundly as the charge-stripe order forms in CsV$_3$Sb$_5$. While **q**$^*_b$ and **q**$^*_c$ are still present and disperse with energy, we no longer observe a dispersive wave vector along the third direction (Fig. 4a,b). Instead in the vicinity we only detect a non-dispersive peak (Fig. 4b, top panel). As **q**$^*_i$ represents the fingerprint of scattering between the small pockets, the change in **q**$^*_a$ is directly tied to the additional renormalization of the hole pockets, which in turn accompanies the emergence of the charge-stripe order. A possible schematic of the charge-stripe order driven modification of the constant energy contour is shown in Fig. 4c, where the size and the shape of the pockets connected by the **Q**$_{4a0}$ wave vector changes. Another intriguing aspect of this band structure change is that **q**$^*_b$ and **q**$^*_c$ now bend away from the high-symmetry Γ-K directions (Fig. 4,f). We note that this is not the case for the equivalent vectors when the $4a_0$ charge order is absent (Supplementary Figure 7). This deviation of **q**$^*_b$ and **q**$^*_c$ from high-symmetry directions demonstrates additional renormalization of the Fermi surface.

The measurements above explored the effects of density waves in the normal state. In the superconducting state of CsV$_3$Sb$_5$, superconducting gap and the coherence peak height also vary spatially in a periodic manner, as reported in a previous experiment [19]. Such modulations are a hallmark of a Cooper pair-density wave phase. The period of the emergent pair-density wave in CsV$_3$Sb$_5$ is about $4a_0/3$ by $4a_0/3$ in-plane, and it coexists with the $4a_0$ charge-stripes and the $2a_0$ by $2a_0$ CDW [19]. Interestingly, the $4a_0/3$ by $4a_0/3$ modulation in real-space translates to about 0.75 **Q**$_{Bragg}$ in reciprocal space, which is exactly the same magnitude and the direction of **q**$^*$ reciprocal-space wave vectors uncovered here. As such, our work also sheds light on the spectroscopic origin of the pair density wave related to the same Fermi pockets (see schematic in Fig. 1d) originally formed by band folding in the $2a_0$ by $2a_0$ CDW state.

The difficulty of pinpointing spectroscopic origins of the $4a_0$ charge-stripe order thus far and its apparent absence in X-ray diffraction experiments prompted a hypothesis that the charge-stripe order may be a surface reconstruction [36]. Our data reveals how the formation of the $4a_0$ charge order accompanies the renormalization of the electronic band structure tied to small hole pockets that are intrinsic parts of the bulk electronic band structure. This in turn highlights charge-stripe ordering as an inherent feature that can be realized in this family of kagome superconductors. It was theoretically proposed that favorably positioned Fermi pockets could drive the emergence of the $4a_0$ charge order [15] (Figure 1d). The combination of our ARPES and STM data suggests that such condition may indeed be satisfied. Further supporting the notion that unidirectional charge ordering can be realized in kagome superconductors, we note that recent scattering measurements unveiled signatures of unidirectional bulk charge correlations in doped $CsV_3Sb_5$ [55]. Interestingly, we also mention that the sole presence of these Fermi pockets is not sufficient to drive the formation of the charge stripe-order in all $AV_3Sb_5$ members, as $\mathbf{q}^*$ is still present in $KV_3Sb_5$ (Fig. 2) although no long-range charge-stripe order is detected in STM measurements [17,28]. It will be of interest to explore if the pair-density wave also emerges in the superconducting state of $KV_3Sb_5$. Future experimental and theoretical work will be necessary to fully understand the physical mechanism necessary to drive these phenomena.

The intriguing renormalization of the Fermi surface in $CsV_3Sb_5$ in the presence of charge-stripe order suggests that a subset of Fermi pockets reconstruct in reciprocal-space. It is conceivable that this in turn may explain some low frequencies in quantum oscillation experiments that are difficult to be captured by a theoretical model that only takes into account the $2a_0 \times 2a_0$ CDW in the kagome plane. We note that the inevitable presence of charge-stripe domains of sub-micron scales [18] hinders the observation of the additional renormalization of pockets by ARPES, as it averages over larger regions of the sample likely spanning stripe domains oriented along all 3 lattice directions. Shubnikov-de Hass [33,38] and de Haas–van Alphen [40] quantum oscillation experiments have detected many low-frequency orbits, several of which carrying a nontrivial Berry phase [38,40]. The pockets observed here are due to Fermi surface reconstruction and could acquire concentrated Berry curvature and orbital magnetic moments if time-reversal symmetry is broken in the CDW state [15]. As a result, these pockets may be tunable by magnetic field, which could be explored in future field-sensitive experiments.

**Acknowledgements**

I.Z. gratefully acknowledges the support from NSF-DMR 2216080 for STM experiments. Work at MIT was supported by the Air Force Office of Scientific Research Young Investigator Program under grant FA9550-19-1-0063, and by the STC Center for Integrated Quantum Materials (National Science Foundation grant no. DMR-1231319). This research used resources of the Advanced Light Source, a US Department of Energy Office of Science User Facility under contract no. DE-AC02-05CH11231. M.K. acknowledges a Samsung Scholarship from the Samsung

Foundation of Culture. Z.W. acknowledges the support of U.S. Department of Energy, Basic Energy Sciences Grant No. DE-FG02-99ER45747 and the Cottrell SEED Award No. 27856 from Research Corporation for Science Advancement. The theoretical calculations were funded, in part, by the Gordon and Betty Moore Foundation EPiQS Initiative, Grant No. GBMF9070 to J.G.C. S.D.W. and B.R.O. gratefully acknowledge support via the UC Santa Barbara NSF Quantum Foundry funded via the Q-AMASE-i program under award DMR-1906325.

## Methods

*Single crystal growth.* Bulk single crystals of $CsV_3Sb_5$ and $KV_3Sb_5$ were grown and characterized as described in Refs. [6,8].

*ARPES measurements.* ARPES experiments were performed at Beamline 7.0.2 of the Advanced Light Source (MAESTRO). The photoelectrons were collected by a hemispherical electron analyzer equipped with a deflector (DA30, Scienta Omicron). $KV_3Sb_5$, $RbV_3Sb_5$, $CsV_3Sb_5$, and Sn-doped $CsV_3Sb_5$ single crystals were cleaved inside an ultra-high-vacuum (UHV) ARPES chamber (~4 × $10^{-11}$ torr). The ARPES data in Fig. 2 were obtained with 97 eV photons ($k_z \approx 0.67\pi/c$) with linear horizontal polarization. The energy and momentum resolutions were better than 20 meV and 0.01 Å$^{-1}$.

*STM measurements.* STM data was acquired using a customized Unisoku USM1300 microscope at the base temperature of 4.5 K and zero magnetic field. We cleave each sample at cryogenic temperature and immediately insert it into the STM head held at 4.5 K [18,28,45]. Spectroscopic measurements were made using a standard lock-in technique with 915 Hz frequency and bias excitation as noted in figure captions. STM tips used were home-made chemically-etched tungsten tips, annealed in a UHV chamber to a bright orange color before STM experiments. We apply the Lawler-Fujita drift-correction algorithm to all our data to align the atomic Bragg peaks onto single pixels with coordinates that are even integers.

*Density functional theory calculations.* Electronic structure calculations were performed using the Vienna ab initio simulation (VASP) package [56,57] based on the Projector augmented wave (PAW) potentials [58,59] and the Perdew-Burke-Ernzerhof (PBE) exchange-correlation functional [60]. Effective tight-binding Hamiltonians projected using Wannier90 code [61] were derived to unfold the band structure in the CDW state [62].

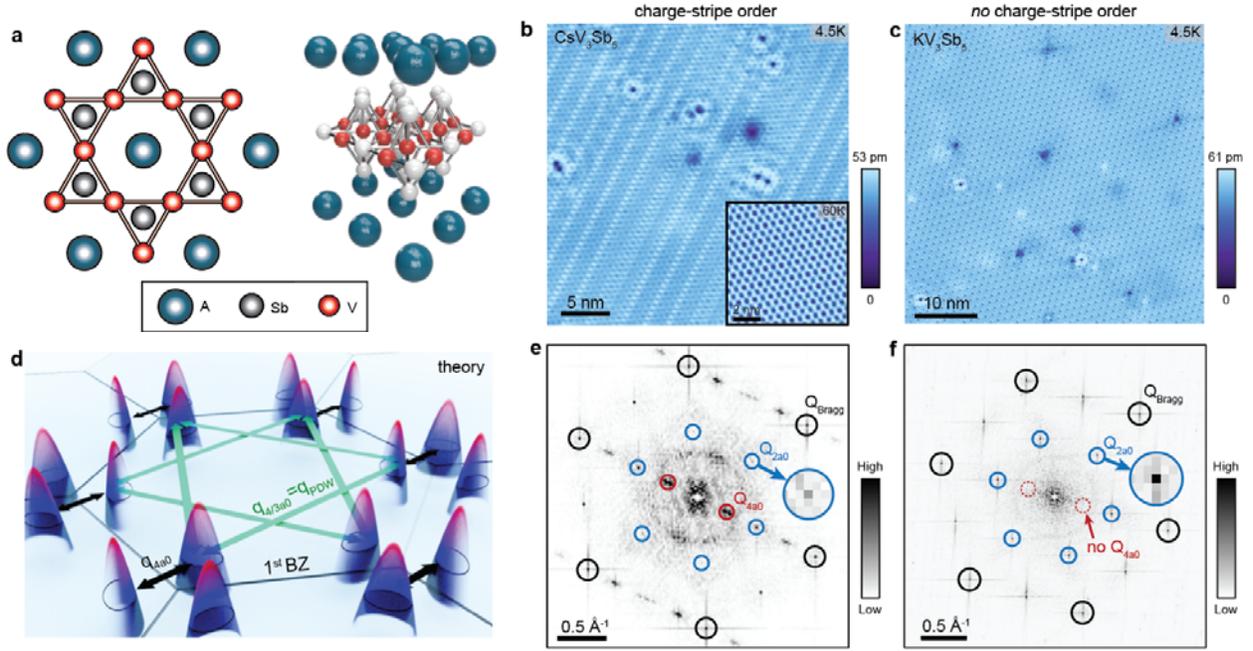

**Figure 1. Crystal structure and emergent charge orders in $AV_3Sb_5$.** (a) A 2D and a 3D ball model of the crystal structure of $AV_3Sb_5$ ($A$=K, Cs, Rb) crystals. STM topographs of (b) $CsV_3Sb_5$ (Sb termination) and (c) $KV_3Sb_5$ (Sb termination) taken at 4.5 K, and (e,f) their associated Fourier Transform (FT). The bottom right inset in (b) is an STM topograph encompassing a smaller region of the Sb termination taken at 60 K, the temperature around which the $4a_0$ charge stripe order disappears. Atomic Bragg peaks $\mathbf{Q}_{Bragg}$, $2a_0$ CDW peaks $\mathbf{Q}_{2a0}$ and charge-stripe order $\mathbf{Q}_{4a0}$ peaks in (c) are circled in black, blue and red, respectively. (d) Schematic of the theoretically expected Fermi pockets and reciprocal space vectors connecting them from Ref. [15]. For simplicity of visualizing relevant pockets in (d), we omit plotting the six additional pockets within the first Brillouin zone, "folded-in" by the $2a_0$ by $2a_0$ CDW. STM setup conditions: $I_{set}$ = 300 pA, $V_{sample}$ = 50 mV; Inset: $I_{set}$ = 30 pA, $V_{sample}$ = 50 mV (b); $I_{set}$ = 400 pA, $V_{sample}$ = 20 mV (c).

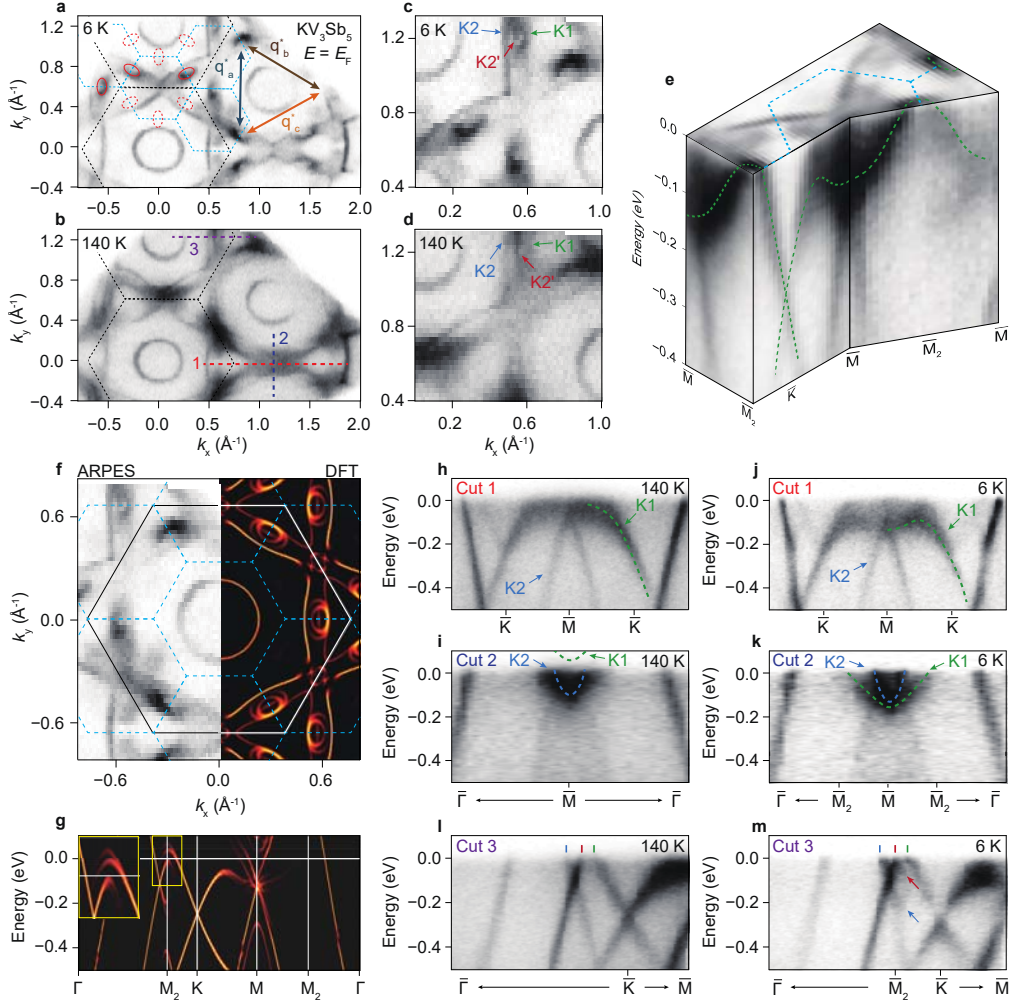

**Figure 2. Momentum-space mapping of Fermi pockets in $KV_3Sb_5$.** Fermi surface of $KV_3Sb_5$ across the first and second Brillouin zone at 6 K (**a**) and 140 K (**b**). Black and blue dashed hexagons represent the Brillouin zone in normal and charge density wave phases, respectively. Red solid and dashed ellipses indicate the Fermi pockets with strong and weak spectral weights, respectively. The three double-sided arrows with gray, brown, and orange show wave vectors $\mathbf{q}^*_a$, $\mathbf{q}^*_b$, and $\mathbf{q}^*_c$, respectively. Fermi surface near $\bar{M}$ point at 6 K (**c**) and 140 K (**d**). Green, blue, and red arrows indicate K1, K2, and K2′ bands, respectively. The K1 band arises from Vanadium $d_{xy}/d_{x^2-y^2}$ orbitals and is associated with a Dirac point at the K point and a *p*-type van Hove singularity (vHS) at the M point near the Fermi level. The K2 and K2' bands arise from Vanadium $d_{xz}/d_{yz}$ orbitals and form a *p*- and an *m*-type vHS, respectively. (**e**) Three-dimensional energy-momentum dispersion near $\bar{M}$ point at 6 K. (**f**) Comparison between experimental and calculated Fermi surface at $k_z = 0.67\pi/c$. The calculated Fermi surfaces assume staggered tri-hexagonal charge order and are three-fold symmetrized to average over three symmetry-equivalent domains. (**g**) Calculated dispersion along high-symmetry points $k_z = 0.67\pi/c$. The inset shows the magnified band dispersion inside the yellow box. Energy-momentum dispersion along $\bar{K}$-$\bar{M}$-$\bar{K}$ (red dashed line in **b**) at 6 K (**j**) and 140 K (**h**), $\bar{\Gamma}$-$\bar{M}$-$\bar{\Gamma}$ (navy dashed line in **b**) at 6 K (**k**) and 140 K (**i**), and $\bar{\Gamma}$-$\bar{M}_2$-$\bar{K}$-$\bar{M}$ (purple dashed line in **b**) at 6 K (**m**) and 140 K (**l**). The K1 and K2 bands (marked with green and blue arrows, respectively) exhibit a hole-like dispersion along the $\bar{K}$-$\bar{M}$-$\bar{K}$ direction (h) and electron-like dispersion along $\bar{\Gamma}$-$\bar{M}$-$\bar{\Gamma}$ (i).

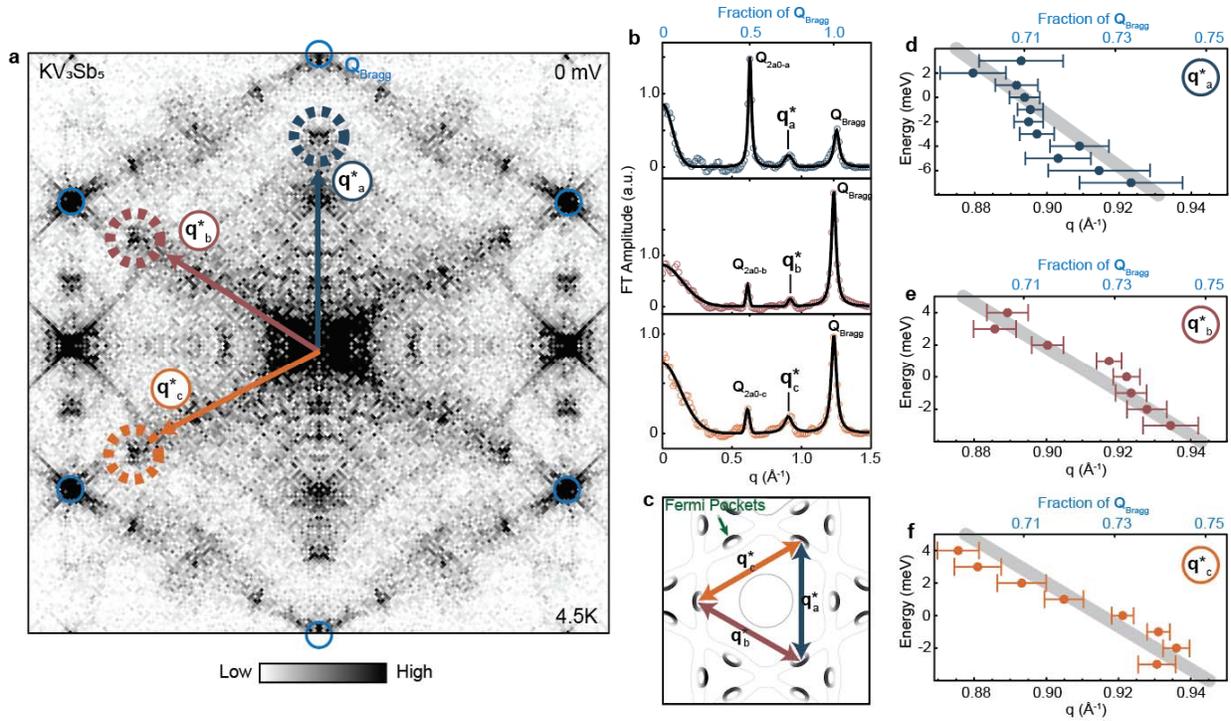

**Figure 3: Dispersive scattering peaks q\* near Fermi level in KV$_3$Sb$_5$.** **(a)** Two-fold symmetrized Fourier transform (FT) of d$I$/d$V$(**r**, $V$= 0 mV) map acquired on the Sb surface of KV$_3$Sb$_5$. Atomic Bragg peaks **Q**$_{Bragg}$ are enclosed in dark blue circles, while the new scattering peaks **q**\*$_i$ ($i$ = $a$, $b$ or $c$) near ¾·**Q**$_{Bragg}$ are dash-circled in gray (**q**\*$_a$), brown (**q**\*$_b$) and orange (**q**\*$_c$) color. **(b)** FT linecuts along the three **Q**$_{Bragg}$ directions (circles) and the Gaussian fittings to each peak in the curve (solid line). The linecut is averaged by 7 pixels perpendicular to the linecut. **(c)** Approximate schematic of the relevant portions of the Fermi surface of KV$_3$Sb$_5$ band structure, with an example of a Fermi pocket called out by the green arrow. The three double-sided arrows in (c) denote scattering processes corresponding to different **q**\* wave vectors in (a). **(d-f)** The position of the **q**\*$_i$ peaks as a function of energy. All three peaks are dispersive with similar band velocities (thick gray line in (d-f) is a visual guide). The position of the peak is extracted from Gaussian fitting as shown for one energy in (b), and the error bars represent the standard error from Gaussian fitting. STM setup conditions: (a) $I_{set}$ = 400 pA, $V_{sample}$ = 20 mV, $V_{exc}$ = 1 mV, 4.5 K.

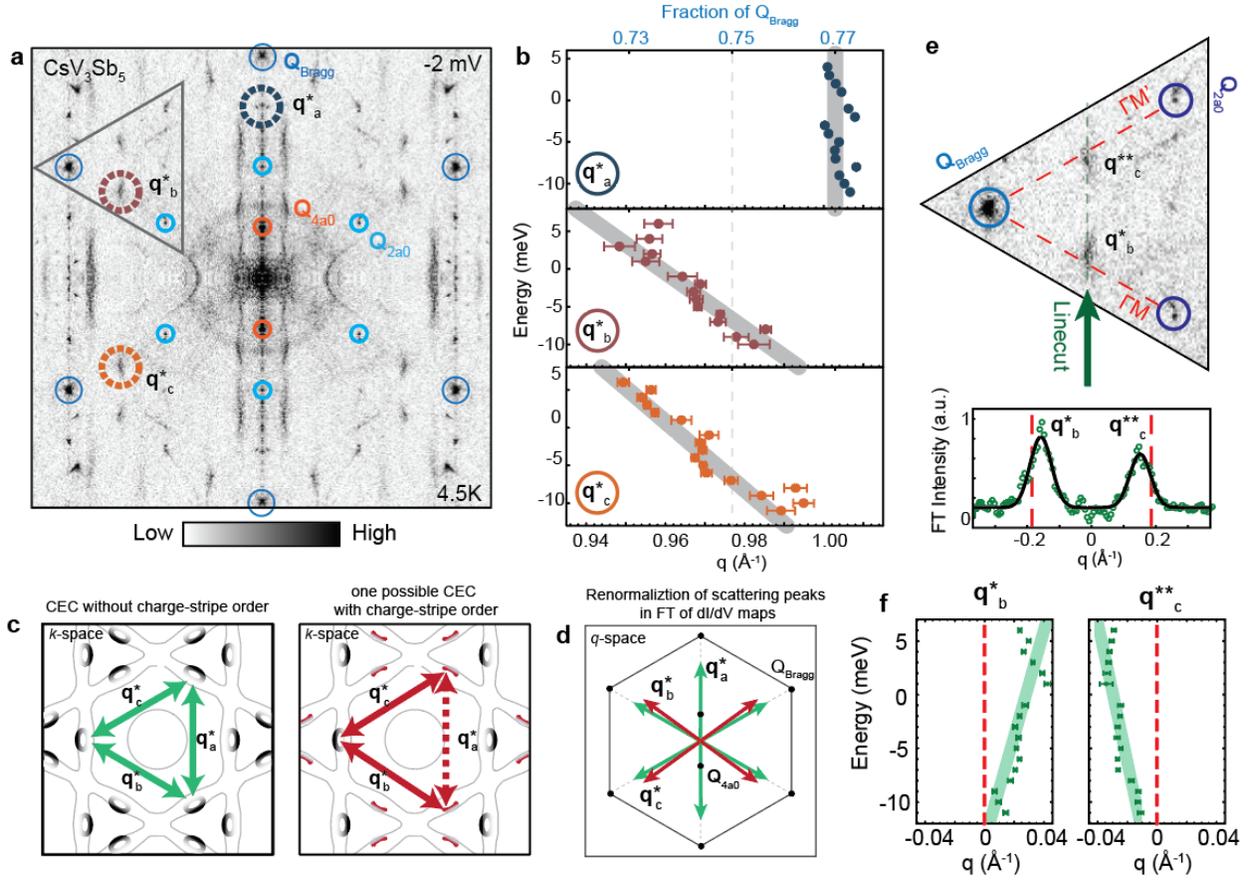

**Figure 4. Renormalization of q* scattering wave vectors in the presence of the charge-stripe order in CsV$_3$Sb$_5$.** (a) A Fourier transform (FT) of d$I$/d$V$(**r**, V = -2 mV) map acquired over a 96 nm square Sb surface of CsV$_3$Sb$_5$. Atomic Bragg peaks **Q**$_{Bragg}$ are enclosed in blue circles, while the new scattering peaks **q**$^*_i$ (i = a, b or c) are dash-circled in dark gray (**q**$^*_a$), brown (**q**$^*_b$) and orange (**q**$^*_c$) color. (b) The dispersion of the **q**$^*_i$ peaks as a function of energy. In contrast to dispersion of **q**$^*_i$ in the absence of charge-stripe order in Figure 3, only **q**$^*_b$ and **q**$^*_c$ are dispersive, while **q**$^*_a$ is static and at a slighly different q-space position. (c) An approximate schematic of the Fermi surface of CsV$_3$Sb$_5$ without (left panel) and with (right panel) charge-stripe order. (d) A comparison of **q**$^*_i$ scattering wave vectors seen in our experiments consistent with the schematics in (c). (e) Zoom on a region in FT space (denoted by a triangle in panel (a)) more clearly showing **q**$^*_b$ and the Bragg reflection of **q**$^*_c$ (labeled **q**$^{**}_c$). Red dashed line connects **Q**$_{Bragg}$ (blue circle) and **Q**$_{2a0}$ (purple circle). The bottom curve in (e) is the linecut (green open circles) taken along the green arrow direction in the top panel, and Gaussian fits (solid black line). The linecut is averaged by 3 pixels along the linecut and 15 pixels perpendicular to the linecut. (f) The positions of the **q**$^*_b$ and **q**$^{**}_c$ in (d) relative to the high symmetry line at each energy for CsV$_3$Sb$_5$ (vertical red dashed line). The position of the peak at each energy is extracted from Gaussian fits as seen in the example in (e). The error bars represent the standard error from Gaussian fitting. STM setup conditions: (a) $I_{set}$ = 40 pA, $V_{sample}$ = -2 mV, $V_{exc}$ = 1 mV, 4.5 K.